\documentclass[aps,pra,english,twocolumn,showpacs,preprintnumbers,amsmath,amssymb,floatfix,superscriptaddress]{revtex4-1}
\usepackage[T1]{fontenc}
\usepackage[latin9]{inputenc}
\usepackage{graphicx}
\usepackage{amssymb}
\usepackage{babel}
\makeatletter

\usepackage[version=3]{mhchem}

\usepackage{color}

\usepackage{ulem}

\makeatother
\usepackage{babel}
\makeatother

\begin{document}

\title{Non-Perturbative Theory for  Dispersion Self-Energy of Atoms}

\author{Priyadarshini Thiyam}
\email{thiyam@kth.se}
\affiliation{Department of Materials Science and Engineering, Royal Institute of Technology, SE-100 44 Stockholm, Sweden}

\author{C. Persson}
\affiliation{Department of Materials Science and Engineering, Royal Institute of Technology, SE-100 44 Stockholm, Sweden}
\affiliation{Department of Physics, University of Oslo, P. O. Box 1048 Blindern, NO-0316 Oslo, Norway}
\affiliation{Centre for Material Science and Nanotechnology, University of Oslo, P. O. Box 1048 Blindern, NO-0316 Oslo, Norway}

\author{I. Brevik}
\affiliation{Department of Energy and Process Engineering, Norwegian University of Science and Technology, NO-7491 Trondheim, Norway}

\author{Bo E. Sernelius}
\email{bos@ifm.liu.se}
\affiliation{Division of Theory and Modeling, Department of Physics, Chemistry and Biology, Link\"{o}ping University, SE-581 83 Link\"{o}ping, Sweden}

\author{Mathias Bostr{\"o}m}
\email{Mathias.Bostrom@smn.uio.no}
\affiliation{Centre for Material Science and Nanotechnology, University of Oslo, P. O. Box 1048 Blindern, NO-0316 Oslo, Norway}

\begin{abstract}
We go beyond the approximate series-expansions used in the dispersion theory of  finite size atoms.  We demonstrate that a correct, and non-perturbative, theory dramatically alters the dispersion self-energies of atoms.  The non-perturbed theory gives as much as 100$\%$ corrections compared to the traditional series expanded theory for the smaller noble gas atoms. 

 \end{abstract}

\pacs{31.30.jh 34.20.Cf 42.50.Lc 03.70.+k}

\maketitle

Until recently most work, using either the quantum electrodynamic (QED) or the semi-classical formalism of dispersion energies,  have relied upon a point dipolar description. Previous work that indeed did incorporate the finite atomic size was based on series expansions, and suggested that the van der Waals interaction contributed to a short range attractive binding energy.\,\cite{Mahanty,Mahanty3}  The aim of a recent work\,\cite{Bos2014} was to demonstrate how keeping the full non-perturbative theory and taking the finite atomic size into account strongly alter the non-retarded van der Waals force and resonance interaction energy at contact distances. Including finite atomic size effects in a non-perturbative theory opens up for the possibility of having van der Waals repulsion when two atoms come very close. Series expansion is a valid approach if the coupling is very weak. 

In the present brief report we present a derivation of the non-perturbative retarded self-energy of atoms and ions in vacuum. Within our new theory, taken without series expansions, we find substantial corrections to the self-energy of an atom.  A useful analytical asymptote is derived and we present some illuminating numerical results  considering the finite size effects for different atoms and ions. 

In the contributions by Mahanty and Ninham\,\cite{Mahanty,Mahanty3} they demonstrated that the secular equation that gives the perturbed eigenmodes of the electromagnetic field due to the presence of a polarizable particle with finite size is
\begin{equation} 
{[\bar I + 4\pi \tilde G(\bar r,\bar r',\xi )] = 0,}
\label{Eq1}
\end{equation} 
where
\begin{equation} 
\begin{array}{l}
\tilde G(\bar r,\bar r',\xi )\\
 = [{\xi ^2}/{c^2}\bar I + {\nabla _r}{\nabla _r}]\int {\tilde G} (\bar r - \bar r',\xi )\alpha (\bar r'' - \bar r',\xi ){d^3}r'',
\end{array}
\label{Eq2}
\end{equation} 
and the free space Green function has the form\,\cite{Mahanty3}

\begin{equation} 
\tilde{G}(\bar{r}-\bar{r'}, \xi)={\frac {\tilde{I}} {(2 \pi)^{3}}} \int d^{3}k {\frac {exp[i \bar{k} (\bar{r}-\bar{r'})]} {(\xi^2/c^2-k^2)}}
\label{Eq3}
\end{equation} 
where $\tilde{I}$ is the unit tensor.

The dispersion self-energy of a finite size isotropic molecule with a Gaussian spread was derived by Mahanty and Ninham\,\cite{Mahanty,Mahanty3}  in a series expanded theory. Here, we present a non-perturbative theory that can be used for atoms and ions in vacuum, and we show that there are often  large corrections  to the results from the approximate series expanded theory. The dispersion self energy is:\,\cite{Mahanty,Mahanty3}
\begin{equation}
{E_s} = \hbar \int\limits_0^\infty  {\frac{{d\xi }}{{2\pi }}} \ln \left[ {\tilde I + 4\pi \tilde G(\bar r ,\bar r ,i\xi )} \right].
\label{Eq4}
\end{equation} 

As was pointed out by Mahanty and Ninham the finite spread of the polarization make the Green function, $ {\tilde{G}(\bar{r},\bar{r},i\xi)}$, convergent (here $\bar{r}$ is the position of the polarization). The polarization cloud of real atoms has a finite spread and we consider as an interesting case a spatial distribution of the atom following an isotropic Gaussian function. This  gives a Green's tensor where the diagonal elements are equal (i.e. $j=x,y$, and $z$ components are equal).\,\cite{Mahanty2,Mahanty3}  The choice for the polarizability tensor is

\begin{equation}
\tilde \alpha (\bar r ,\xi ) = \tilde I({\pi ^{ - 3/2}}{a^{ - 3}}) \times {e^{ - {r ^2}/{a^2}}}\alpha (\xi ),
\label{Eq5}
\end{equation}
where $a$ is the gaussian radius.
The fully retarded expression can be shown to be,
\begin{equation}
{G_j^{Ret}(i\xi ) =\frac{{\alpha (i\xi )}}{{3{{(2\pi )}^3}}} I(i\xi )  ,}
\label{Eq6}
\end{equation}
where

\begin{equation}
\begin{array}{l}
I(i\xi ) = \frac{{{\pi ^{3/2}}}}{{{{(a/2)}^3}}}\left( {1 + \frac{{{\xi ^2}{a^2}}}{{{c^2}}}} \right)\\
\quad \quad \quad  - 4{\pi ^2}\frac{{{\xi ^3}}}{{{c^3}}}{e^{{{(\xi a/2c)}^2}}}\left[ {1 - erf\left( {\frac{{\xi a}}{{2c}}} \right)} \right].
\end{array}
\label{Eq7}
\end{equation}

In the non-retarded limit this is reduced to
\begin{equation}
{G_j^{NR}(i\xi ) = [\alpha (i\xi )]/[3{\pi ^{3/2}}{a^3}].}
\label{Eq8}
\end{equation}

The traditional way to treat these integrands is to make a series expansion of the logarithm in Eq. (4) and keep only the lowest order term, i.e. $\ln (1 + x) \approx x$. However, the energy from the eigenmodes for an isotropic atom in vacuum is given by the sum of the equal $j=x$, $y$, and $z$ contributions from the secular equation.  We find:

\begin{equation}
{E_s} = \sum\limits_{j = x,y,z} \hbar  \int\limits_0^\infty  {\frac{{d\xi }}{{2\pi }}} \ln \left[ {1 + 4\pi {G_j}(i\xi )} \right].
\label{Eq9}
\end{equation} 

Mahanty demonstrated that when the retarded Green's function is substituted into the series expanded expression, the main contribution in the integration must come from the characteristic absorption frequencies of the atomic system. For the corresponding values of $a \xi/2c \approx a/2 \lambda_{c}$ , where $\lambda_{c}$ is the wavelength of a characteristic absorption line the retardation effects are negligible. In the retarded limit one must use a cut-off as proposed by Mahanty in similar calculations for the Lamb shift.\,\cite{Mahanty}  However, we focus on the non-retarded limit where there is no need for a cut-off and we obtain the following  expression,
\begin{equation}
{E_s^{NR}} = 3\hbar \int\limits_0^\infty  {\frac{{d\xi }}{{2\pi }}} \ln \left[ {1 + \frac{{4\pi }}{{3{\pi ^{3/2}}{a^3}}}\alpha (i\xi )} \right].
\label{Eq10}
\end{equation} 
To illustrate the point we use a simple one oscillator model for the atomic polarizability, $\alpha (i\xi)=\alpha(0)/(1+\xi^2/\omega_0^2)$. We find that a non-perturbative theory gives the following expression for the dispersion self-energy of isotropic polarizable particles (with a Gaussian polarization spread) in vacuum  

\begin{equation}
{E_s^{NR}} = \frac{{3\hbar {\omega _0}}}{2}[ - 1 + \sqrt {1 + 4\alpha (0)/(3{a^3}\sqrt \pi  )} ].
\label{Eq11}
\end{equation}

\begin{table}
\caption{The finite size dispersion self-energy, $E_{\rm{s}}$, for noble gas atoms and two ions. The subscript expanded indicates that the result is from using series expansion of the logarithm in the integrand. All energies are in eV. The input data were taken from Refs. \cite{atompaper,Mahan,ParsonsA,ParsonsB}.}
\begin{tabular}{cccccc}
\colrule
\colrule
\multicolumn{1}{c}{Element}   &\multicolumn{1}{c}{$E_{\rm{s}}^{{\rm{NR,full}}}$} &\multicolumn{1}{c}{$E_{\rm{s}}^{{\rm{NR,expanded}}}$}  \\
\colrule
He&71.2 &131.5 \\
Ne&104.9 &220.3 \\
Ar&37.6 &62.1 \\
Kr&29.9 &47.5  \\
$Na^{+}$&20.2 &22.4 \\
$Cl^{-}$&5.2 &6.0 \\
\colrule
\colrule
\end{tabular}
\label{values}
\end{table}

We give in Table I the results found when using our  non-perturbative theory and as comparison also the results from using the approximate series expanded theory.  As input we have used atomic radii and static polarizabilities given by Hohm and Thakkar\,\cite{atompaper}  and characteristic frequencies given by Mahan and Subbaswamy\,\cite{Mahan}. For the ions we use the input data from Parsons and Ninham.\,\cite{ParsonsA,ParsonsB} 

The effects of a non-perturbative theory ought in principle to be detectable experimentally, although it is a challenge to measure self-energies directly.  What is practically possible, is to verify our results indirectly in experiment.  Solvation energies of atoms and ions in a dielectric medium (i.e. changes of the self-energy in a vacuum compared to in a medium) can be measured\,\cite{Honig,Bost2004,Tim}. Latimer et al\,\cite{Latimer} were able to fit experimental solvation energies (or rather the related heats of solvation) to the Born equation by increasing the effective radius of the ions.  Self-energy changes have also been shown to influence permeabilities across dielectric membranes \,\cite{Pars1969,Bost2005}. The difference between perturbative and non-perturbative theories will be much reduced in a dielectric medium due to the factor $1/\epsilon$ (where $\epsilon$ is the frequency dependent dielectric function of a medium). Hence the results obtained for instance for the permeability of atoms across a membrane will be changed mainly due to changes in the self-energy in vacuum.  A factor of 2 difference for these self-energies compared to those  obtained from a  series-expanded theory ought  therefore  to be measurable with existing experimental equipment for solvation free energies and permeabilities. 
Large corrections have been overlooked in the past when performing series expansions of the logarithmic terms before including the finite atomic size.\,\cite{Mahanty2} The self-energy is also known to give a contribution to the Lamb shift.\,\cite{Mahanty} 
When series expanding Eq.\,(\ref{Eq10}), we rederive the dispersion self-energy found by Mahanty.\cite{Mahanty} However, the validity of a series expansion assumes that $\alpha(0)/a^3$ is much smaller than unity which is not always the case.


MB and CP acknowledge support from the Research Council of Norway (Project: 221469).  CP  acknowledges support from  the Swedish Reseach Council (Contract No. C0485101).  PT acknowledges support from the European Commission.

\end{document}